\documentstyle[aps,preprint,epsfig]{revtex}

\title{Tile Hamiltonian for Decagonal AlCoCu}

\author{Ibrahim Al-Lehyani$^{a,b}$ and Mike Widom$^{a}$}

\address{$^{a}$Department of Physics, Carnegie Mellon University,
Pittsburgh, Pennsylvania 15213}
\address{$^{b}$Department of Physics, King AbdulAziz University, Jeddah, Saudi Arabia}
\begin{document}

\maketitle
\date{\today}

\begin{abstract}
A tile Hamiltonian (TH) replaces the actual atomic interactions in a quasicrystal with effective interactions between and within tiles. We studied Al-Co-Cu decagonal quasicrystals described as decorated Hexagon-Boat-Star (HBS) tiles using {\em ab-initio} methods. The dominant term in the TH counts the number of H, B and S tiles. Phason flips that replace an HS pair with a BB pair lower the energy. In Penrose tilings, quasiperiodicity is forced by arrow matching rules on tile edges. The edge arrow orientation in our model of AlCoCu is due to Co/Cu chemical ordering. Tile edges meet in vertices with 72$^\circ$ or 144$^\circ$ angles. We find strong interactions between edge orientations at 72$^\circ$ vertices that force a type of matching rule. Interactions at 144$^\circ$ vertices are somewhat weaker.
\end{abstract}
\maketitle
\newpage

\section{Introduction}
\label{sec:intro}
Both quasicrystals and ordinary crystals are made of elementary building blocks. In crystals, copies of a single building block (known as a unit cell) are arranged side by side to cover the space periodically. In quasicrystals, building blocks are arranged to cover the space quasiperiodically. Two approaches to build quasilattices have been proposed. One approach uses a single unit cell, but allows adjacent cells to overlap. Gummelt~\cite{Gummelt} proved that using a limited number of overlapping positions between decagons produces a quasicrystal structure. Steinhardt and Jeong~\cite{SteinhardtJoeng} further proved that the overlapping conditions can be relaxed when supplemented by the maximization of a specific cluster density to produce quasilattices. This is the ``quasi-unit cell'' approach. 

In the other ``tiling'' approach, space is covered  with building blocks called tiles. Two or more tile types are used~\cite{patterns}. No overlapping is allowed, and depending on the way the tiles are arranged, quasicrystal structures can be produced. Matching rules between tiles govern the local tile configurations by allowing only a subset of all possible arrangements. Globally, the matching rules enforce quasiperiodicity~\cite{Penrose,Gardner}. Penrose proposed his famous matching rules before the discovery of quasicrystalline materials. In 2D, Penrose tiles are fat and thin rhombi (Fig~\ref{fig:penrose}a). Edges are assigned arrow decorations which must match for common edges in adjacent tiles. In perfect quasicrystals these rules are obeyed everywhere. The very restrictive Penrose matching rules are sufficient, but not necessary, to force quasiperiodicity. Socolar~\cite{Socolar} showed that weaker matching rules can still force quasilattices. The less restrictive set of rules are derived by allowing bounded fluctuations in perp space. Furthermore, matching rules can be abandoned entirely and quasiperiodicity may arise spontaneously in the most probable random tiling~\cite{henley1}.

A fundamental question is whether matching rules are enforced by energetics of real materials. Burkov~\cite{Burkove93} proposed matching rule enforcement by chemical ordering of Co and Cu among certain cites in Al-Co-Cu. However, that model involved an unnatural symmetry linking Co sites to Cu sites. Cockayne and Widom~\cite{CWmodel} deduced a different, physically realistic, type of Co/Cu ordering based on total energy calculations. In their model, tile edges are assigned arrow direction based on their Co/Cu decorations (Fig.\ref{fig:penrose}b). The suggested physical origin of Co/Cu chemical ordering rests on the status of Cu as a Noble Metal with completely filled d orbitals, unlike normal transition metals such as Co.

In a tiling model of quasicrystals, the actual atomic interactions in the system Hamiltonian can be replaced with effective interactions between and within tiles~\cite{tileham}. The resulting tile Hamiltonian is a rearrangement of contributions to the actual total energy. In simple atomic interaction pictures (pair potentials for example) the relation between the two (actual atomic interactions and tile Hamiltonian) is straight forward. It might be difficult to find the relations between them for more complicated atomic interactions (many body potentials, or full {\em ab-initio} energetics, for example) but it is theoretically possible. The tile Hamiltonian includes terms which depends only on the number of tiles and other terms for different interactions. The tile Hamiltonian greatly simplifies our understanding of the relationship between structure and energy, and it is a reasonable way to describe the tiles.

Space can be tiled in many different ways, even when holding the number of atoms or the number of similar tiles fixed. Figure~\ref{fig:app132} shows three different tiling configurations of 132 atoms. The first two have the same tiles arranged differently. The third has the same atoms but different tiles. These are called quasicrystal approximants (crystals that are very close to quasicrystals in structure and properties). One advantage of approximants is that they can be studied using conventional tools developed for ordinary crystals. 

In a previous paper~\cite{ICQ7}~we studied matching rules in decagonal Al-Co-Cu using a limited group of quasicrystal approximants. Some specific details of the tile Hamiltonian couldn't be extracted from our limited data set. Here we study more thoroughly the set of rules controlling these compounds, using different techniques and a much bigger set of approximants.

We describe our model of decagonal Al-Co-Cu in section~\ref{sec:decagonalalcocu} of this paper. Section~\ref{sec:abinitiostudy} gives our detailed calculations using {\em ab-initio} methods. We extract a set of parameters that allow an excellent approximation to the total energy. Similar calculations done using pair potentials are described in section~\ref{sec:pairpotentials} for comparison. In section~\ref{sec:othereffects}, we talk about various other effects that could be considered in a more accurate model. We analyze our findings and study their implications for Al-Co-Cu compounds in section~\ref{sec:dis}.

\section{Decagonal Al-Co-Cu Model}
\label{sec:decagonalalcocu}
Decagonal Al-Co-Cu quasicrystals have been studied by many authors, theoretically~\cite{Burkove93,CWmodel,alconitheory} and experimentally~\cite{alconiexp}. Cockayne and Widom~\cite{CWmodel} employed mock-ternary pair potentials to propose a model based on tiling of space by Hexagon, Boat and Star tiles (HBS) decorated deterministically with atoms (Fig.~\ref{fig:app132}). The tile edge length is 6.38~\AA. Tile vertices are occupied by 11-atom clusters. Each cluster consists of two pentagons of atoms stacked on top of each other at $\frac{1}{2}c$=2.07~\AA~and rotated by 36$^\circ$. The pentagon in one layer contains only Al atoms. The pentagon in the other layer contains a mixture of transition metal (TM) atom species and can contain also Al atoms. The mixed Al/TM pentagon contains an additional ``vertex'' Al atom at its center. All TM atoms surrounding a vertex Al atom belong to tile edges. Decagonal clusters meet along pairs of Co/Cu atoms. It was shown in the original model~\cite{CWmodel} that TM atoms prefer to alternate in chemical species on tile edges. This was confirmed later by {\em ab-initio} calculations using the Locally Self-consistent Multiple Scattering (LSMS) method~\cite{ICQ7,LSMS}, and is confirmed again in this study.

We assign arrows to edges based on their TM atom decorations. By our definition, an arrow points from the Cu atom towards the Co atom. Tile edges meet in vertices of 72$^\circ$ or 144$^\circ$ angles. An angle is of type ``i'' if both edges point in towards their common vertex. Types ``o'' and ``m'' are out- and mixed-pointing, respectively.

\begin{figure}
\centerline{\epsfig{figure=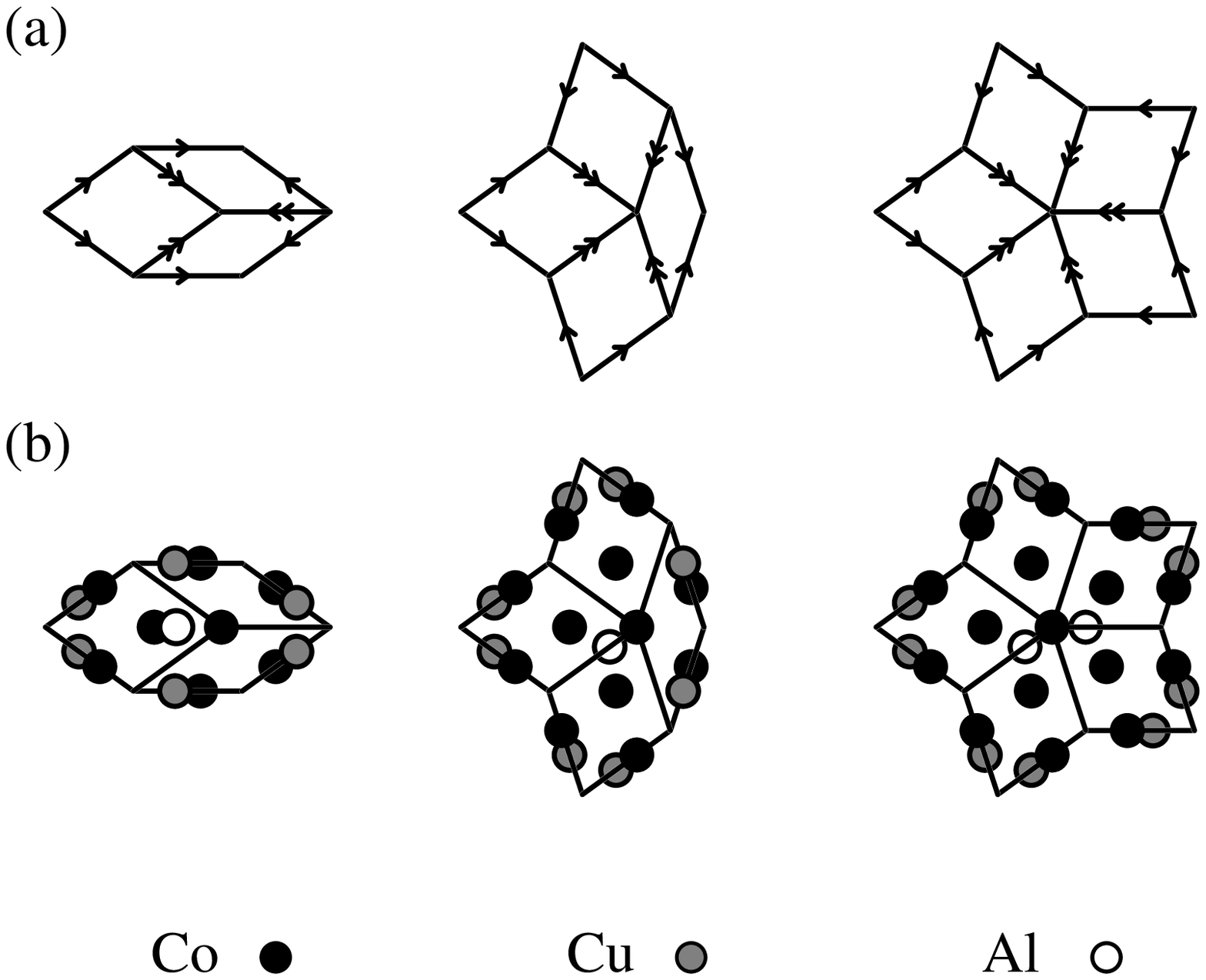}}
\vspace{20pt}
\caption{HBS tiles and their decompositions to Penrose tiles (a) and atomic decorations (b). In (b), only TM and symmetry breaking Al atoms are shown.}
\label{fig:penrose}
\end{figure}

The HBS tiles are composed of Penrose rhombi with double arrow matching rules satisfied (by definition) inside the HBS tiles. Some of their properties are summarized in table~\ref{tab:tiles}. Quasiperiodic tilings can then  be constructed from HBS tiles obeying the single-arrow matching rules (Fig.~\ref{fig:penrose}).

We choose to define a tile Hamiltonian for the system
\begin{equation}
\label{eq:tileham}
H=\sum_{\alpha}n_{t}^{\alpha}E_{t}^{\alpha}+\lambda_{72}\sum_{\beta}n_{72}^{\beta}E_{72}^{\beta}+\lambda_{144}\sum_{\beta}n_{144}^{\beta}E_{144}^{\beta}.
\end{equation}
Here $n_{t}^{\alpha}$ is the number of specific tile type $\alpha$ which can be H, B or S; $n_{72}^{\beta}$ and $n_{144}^{\beta}$ are the numbers of 72$^\circ$ and 144$^\circ$ angles and $\beta$ defines the angle type which can be i, o or m. We fit energy parameters in the Hamiltonian (\ref{eq:tileham}) to achieve $H \approx E_{tot}$ for a wide class of structures. The tile energies are more important than the other terms, and it turns out that the 72$^\circ$ angle interactions are more important than the 144$^\circ$'s. It is desirable to check how each term can alter the system energies. We set $\lambda_{72}$ and $\lambda_{144}$ to 1 or 0 for the purpose of including and excluding the 72$^\circ$ and 144$^\circ$ angle interactions.

An H tile contains 25 atoms, counting those on the perimeter fractionally. All but 3 internal atoms (one Al and two Co) belong to vertex decagonal clusters. The internal Co atoms occupy symmetric sites, but the internal Al atom breaks the symmetry by residing on one of two geometrically equivalent positions between the two Co atoms (Fig.~\ref{fig:penrose}b, left). Its interaction with tile edge TM atoms determines the preferred position. A B tile has 41 atoms. An internal Al atom breaks the symmetry (Fig.~\ref{fig:penrose}b, center) by residing in one of two equivalent sites. An S tile has 57 atoms. Two internal Al atoms break the symmetry (Fig.~\ref{fig:penrose}b, right) by occupying any two of five equivalent sites as long they are 144$^\circ$ apart. A phason flip can switch a specially arranged star and hexagon into a pair of boats. This is shown in Fig.~\ref{fig:app132}c outlined by a dashed line (compare with Fig.~\ref{fig:app132}b). 

The present structure differs slightly from the original model~\cite{CWmodel}. In the original model, Cu atoms take certain symmetry-breaking positions inside the boat and the star which makes it difficult to parameterize their interactions. We choose to replace the Cu atoms in the tile interiors with Al atoms.  Also in the original model, symmetry-breaking Al atoms inside H, B and S were placed in averaged sites between two Co atoms. Their vertical heights lay midway between the two main atomic layers. Here we place them in the main atomic layers as shown in Fig.~\ref{fig:penrose}b. In terms of the atomic surfaces~\cite{CWmodel}, the atomic surface (AS2) that is mainly Al with a thin ring of Cu becomes pure Al, and the Al atomic surface between layers (AS3) fills the hole in the pure Al atomic surface (AS2).

\begin{figure}
\centerline{\epsfig{figure=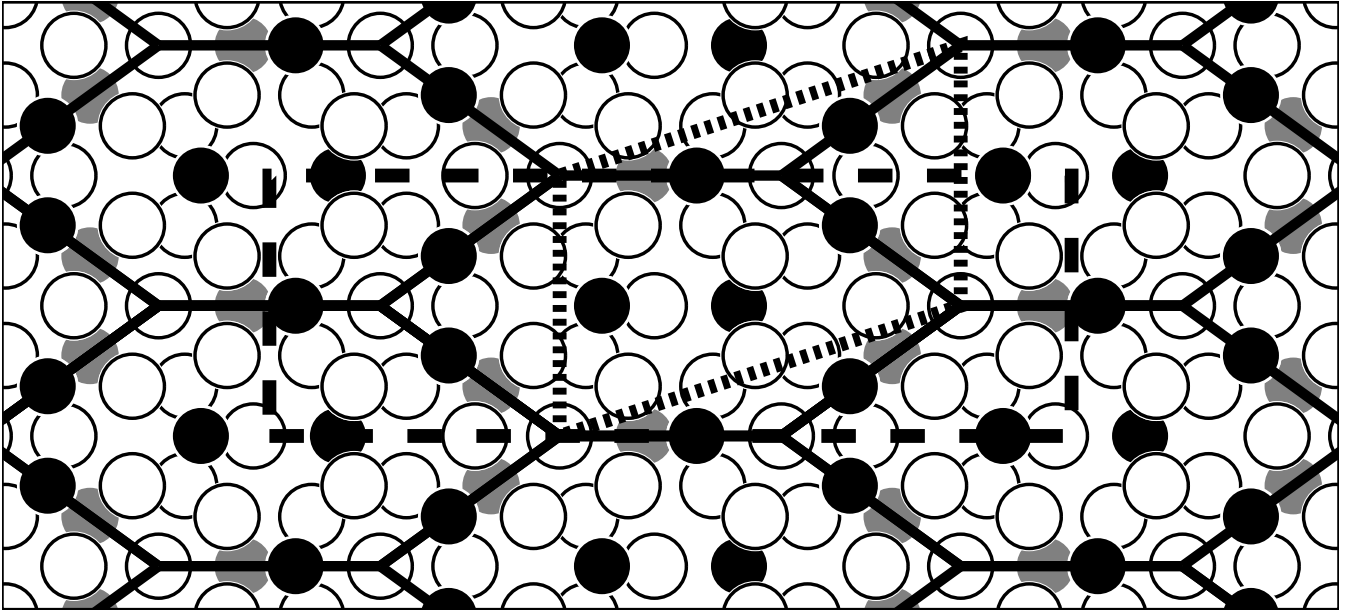}}
\vspace{20pt}
\caption{Filling space with hexagons. This approximant has one hexagon per monoclinic cell (fine dashing, H$_1$ structure in the text). The unit cell has 25 atoms. The cell can be doubled to get a 50-atom orthorhombic unit cell (coarse dashing, H$_2$ structure in the text).}
\label{fig:h2_mono}
\end{figure}

\begin{figure}
\centerline{\epsfig{figure=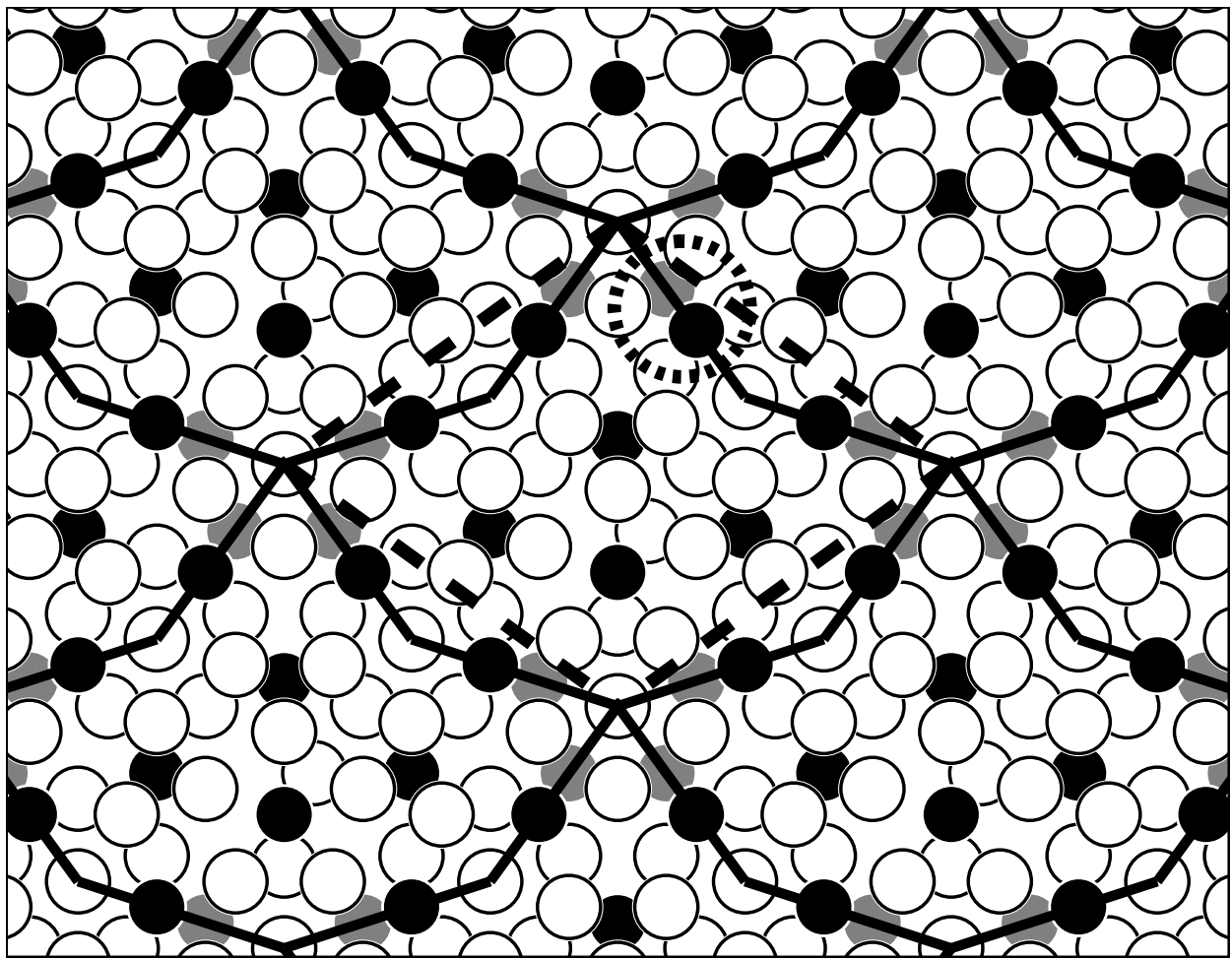}}
\vspace{20pt}
\caption{Single-boat (B$_1$) approximant containing 41 atoms per monoclinic cell. The circled CoCu pair was used for our convergence study.}
\label{fig:b1}
\end{figure}

\begin{figure}
\centerline{\epsfig{figure=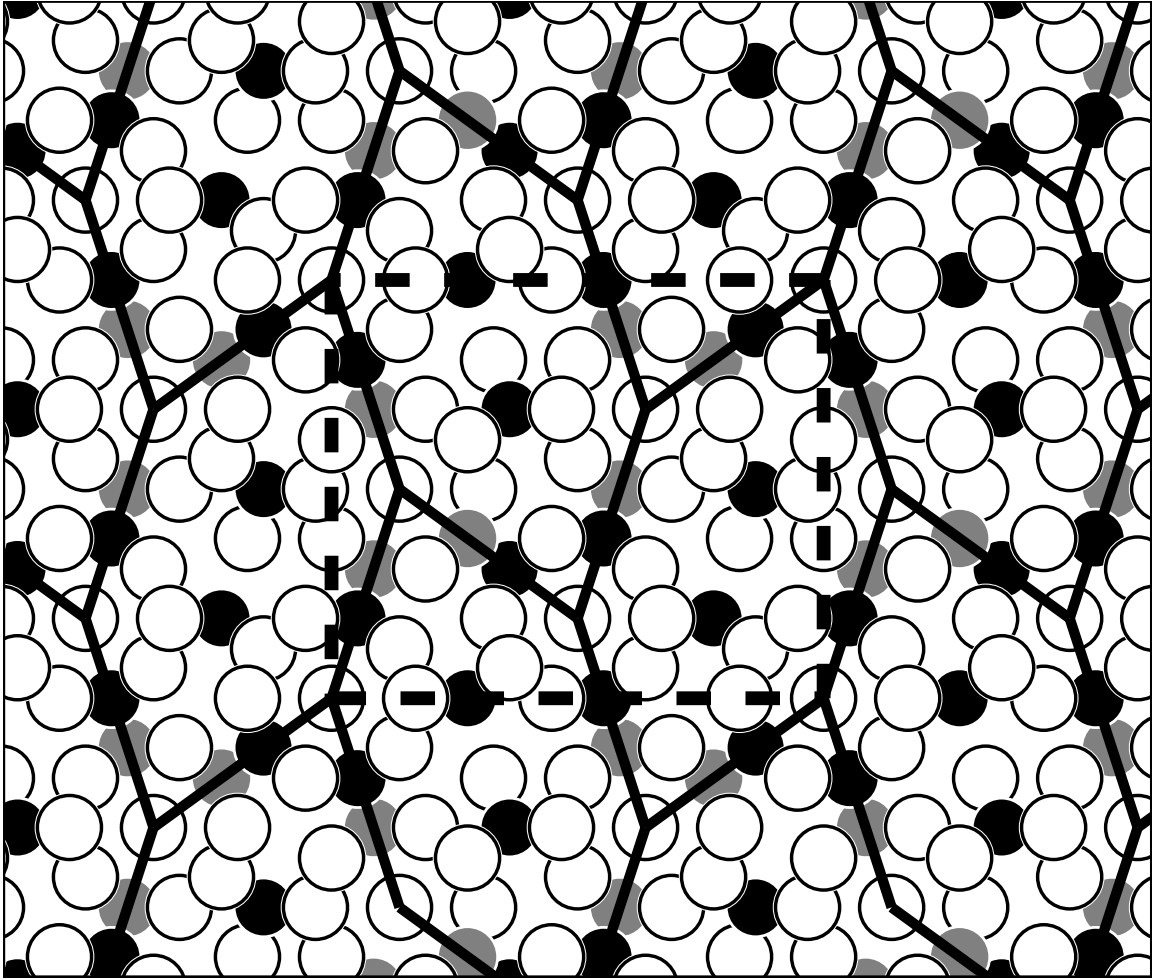}}
\vspace{20pt}
\caption{Orthorhombic two-hexagon approximant (H$_2'$) with a different arrangement of hexagons from fig~\ref{fig:h2_mono}.}
\label{fig:h2_ortho}
\end{figure}

Many different quasicrystal approximants are exploited here to study different terms in the tile Hamiltonian. All the approximant tilings we used are listed in table~\ref{tab:structures} with some of their properties.  The smallest tiling is the monoclinic single-hexagon approximant H$_1$ (Fig.~\ref{fig:h2_mono}). It contains one ``horizontal'' and two ``inclined'' tile edges. For the decoration shown, equation~(\ref{eq:tileham}) becomes $H=E_{t}^{H}+\lambda_{72}(E_{72}^{i}+E_{72}^{o})+\lambda_{144}(2E_{144}^{i}+2E_{144}^{o})$. The next bigger approximant is a 41-atom single-boat B$_1$ (Fig.~\ref{fig:b1}) for which $H=E_{t}^{B}+\lambda_{72}(3E_{72}^{o})+\lambda_{144}(2E_{144}^{i}+E_{144}^{o})$ when decorated as shown. Stars alone do not tile the plane, so a single-star unit cell approximant is not possible. Two-hexagon approximants can be constructed in orthorhombic cells, either by a genuine orthorhombic structure H$_2'$ (Fig.~\ref{fig:h2_ortho}) or by doubling the monoclinic H$_1$ cell to create H$_2$ shown in Fig.~\ref{fig:h2_mono}. In each case, $H=E_{t}^{H}+\lambda_{72}(2E_{72}^{i}+2E_{72}^{o})+\lambda_{144}(4E_{144}^{i}+4E_{144}^{o})$. The doubling process gives more freedom in controlling edge arrow orientations by performing Co/Cu swaps. Similarly, a two-boat approximant can be constructed, either by doubling the single boat tiling B$_1$ or by tiling the two boats as shown in B$_2$ (Fig.~\ref{fig:b2}), with $H=E_{t}^{B}+\lambda_{72}(3E_{72}^{o}+3E_{72}^{m})+\lambda_{144}(3E_{144}^{i}+E_{144}^{m}+2E_{144}^{o})$ when decorated as shown.

To isolate the tile Hamiltonian parameters $E_{t}^{\alpha}$ and $E_{\theta}^{\beta}$, even larger approximants are needed. The three approximants in Fig.~\ref{fig:app132} each have 132 atoms per unit cell but different tile configurations. Two of them (B$_2$H$_2'$ and S$_1$H$_3$ in Fig.~\ref{fig:app132}b and c respectively) are related to each other by a single phason flip. The phason flip turns out to raise the energy, indicating that stars are disfavored in these compounds. Angle orientations are investigated by swapping a TM pair on tile edges, reversing the directions of the edge arrows.

Long range interactions and other small terms omitted from  the tile Hamiltonian (\ref{eq:tileham}) can be estimated by swapping pairs surrounded by symmetric CoCu bonds on their sides. For example, the CoCu bonds on the sides of the horizontal tile edges of H$_2$ can be specially arranged to cancel all angle interactions $E_{\theta}^{\beta}$ and leave only other effects.

\section{{\em Ab-initio} Study}
\label{sec:abinitiostudy}
For our calculations we employ {\em ab-initio} pseudopotential calculations utilizing the Vienna {\em Ab-initio} Simulation Package (VASP) program~\cite{VASP}. We use ultrasoft Vanderbilt type pseudopotentials~\cite{vander} as supplied by G. Kresse and J. Hafner~\cite{VASPpot}. Our calculations are carried out on the Cray T3E and on the newly installed Compaq TCS machine at the Pittsburgh Supercomputer Center. 

The k-space mesh size (among other parameters) determines the accuracy of the calculations. Bigger k-space grids are more accurate but more expensive in calculation time. One has to find a balance between  the number of atoms in a unit cell and the size of the k-space grid in order to fit within the available computer resources. As explained before, we use many different approximants for our study each with its own convergence behavior. The k-space mesh is increased until a consistency of about 0.02 eV is reached in worst cases. But within a reasonable use of our allocated computer times we are able to get better convergence (0.002-0.01 eV) for most structures.

Convergence test calculations are summarized in table~\ref{tab:convergenceb1} for our B$_1$ tiling and in table~\ref{tab:convergenceb2h2-1} for our   B$_2$H$_2$ tilings. We compared two slightly different structures for each tiling, differing in orientation of a single arrow (CoCu pair circled in Fig.~\ref{fig:b1} and Fig.~\ref{fig:app132}a). The tables suggest that at the chosen k-space grid used in our calculations, marked by a star in the tables, the accuracy is better than 0.02 eV. The smallest grid of 1x1x1 (single k-point in the center of the k-space unit cell) for our B$_1$ approximant takes about 8 minutes on the T3E machine (450 MHz alpha processors) using 8 processors. The largest grid we use for B$_1$ of 5x5x15 (188 independent k-point) takes 4.8 hours on 64 processors.  The   B$_2$H$_2$ structure has 132 atoms per unit cell. The smallest 1x1x1 grid takes 2 hours on 8 processors. The largest grid of 2x2x10 (20 independent k-points) takes about 6 hours on 64 processors. 

For a fixed number of processors, calculation time grows linearly with the number of independent k-points. Calculation time decreases linearly with increasing number of processors only up to about 16 processors. Beyond that, the total charging time (number of processors$\times$elapsed time) increases notably. Large structures and big k-space grids require large numbers of processors because they need more memory.

Note from tables~\ref{tab:convergenceb1} and~\ref{tab:convergenceb2h2-1} that the more isotropic the distribution of the k-space grid points along k$_x$, k$_y$ and k$_z$, the faster the  convergence.  For our B$_1$ approximant, meshes that most isotropically  distribute k-space points are 1x1x3 and its multiples, but for finer meshes 4x4x11 is slightly more isotropic than 4x4x12. For fixed numbers of k$_x$ and k$_y$ points, the total energy converges towards its limiting value as the number of k$_z$ points approaches its isotropic value. In all our structures, we choose the most isotropic distribution of k-space points possible.

Many rearrangements of edges are performed by swapping TM atom pairs and the different structure energies are calculated. Most of these are done for the large approximants because they give more configurational freedom. In general, pure 72$^\circ$ angle interaction parameters ($E_{72}^{\beta}$) can be deduced if the bond to be swapped has an equal number of 144$^\circ$ connections on each side (since we can arrange for 144$^\circ$ interaction to cancel) and either a single 72$^\circ$ connection which can be attached to any side or two 72$^\circ$ connections attached to the same side. An example is outlined by a dashed rectangle in Fig.~\ref{fig:app132}a. The bond is surrounded by a single 144$^\circ$ on each side. Both point towards the middle bond, making E$_{144}^i$ on one side, and E$_{144}^m$ on the other. One extra 72$^\circ$ on one side makes E$_{72}^o$. If the middle pair is swapped, the 144$^\circ$ angle interactions will be conserved. The swap merely switches their sides, while the E$_{72}^o$ becomes E$_{72}^m$. The difference in energy  between the swapped and the basic structures yields the difference $E_{72}^m-E_{72}^o=E_{after}-E_{before}=0.26$ eV. Reversing the 72$^\circ$ outer bond and then making the swap again can give $E_{72}^i-E_{72}^m=0.35$ eV. For pure 144$^\circ$ interactions, we use a middle bond surrounded by two 144$^\circ$ angle on one side and a single 144$^\circ$ angle on the other side. We arrange for any 72$^\circ$ angles to cancel. The pair surrounded by a dashed circle in Fig.~\ref{fig:app132}a is one example that yields $E_{144}^m-E_{144}^o=0.06$ eV. 

\begin{figure}
\centerline{\epsfig{figure=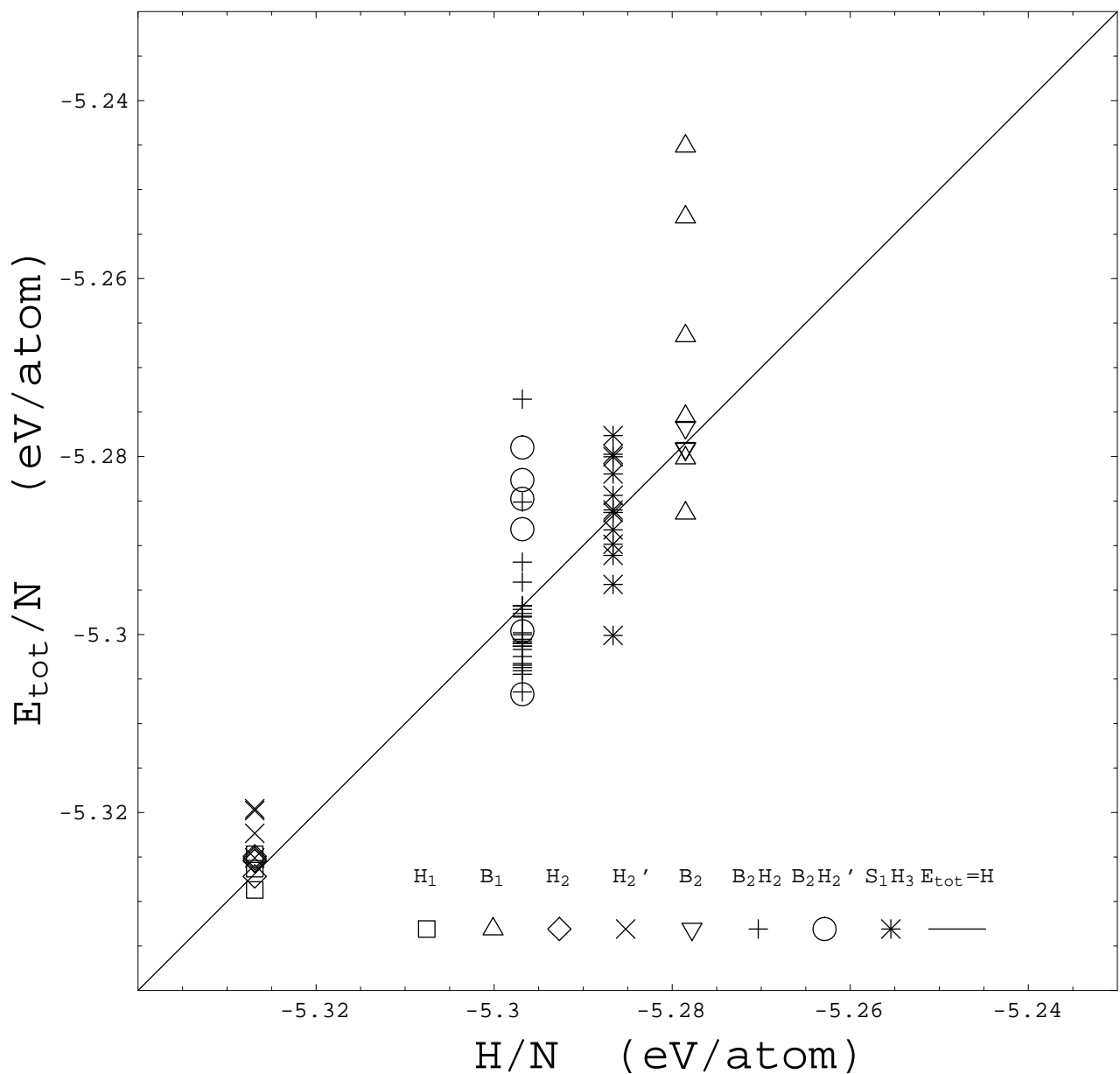,angle=0}}
\vspace{20pt}
\caption{Plots of calculated structure energies vs. our model expectations using only tile energies (turning off angle interactions). The spread of energies vertically is due to angle interactions not accounted for in the model energy when $\lambda_{72}=\lambda_{144}=0$. The diagonal line indicates $H=E_{tot}$.}
\label{fig:vasp3}
\end{figure}

We compute energies for an over-complete set of structures and use a least square fit to determine average parameter values. First, we fit with $\lambda_{72} = \lambda_{144} = 0$ in equation (\ref{eq:tileham}), leaving only three adjustable parameters $E_t^H$, $E_t^B$ and $E_t^S$. The fit finds the values of our parameters that minimize the root mean square of the difference between the calculated $E_{tot}$  and model $H$ values for an ensemble of tilings with different angle orientations. The fitted values $E_t^{B}, E_t^{H} \mbox{ and }E_t^{S}$ are shown in the second column of table~\ref{tab:fitting}. The fitting is shown in Fig.~\ref{fig:vasp3}. The graph shows clearly that the three-parameter fit is not adequate because there is a significant variations of $E_{tot}$ among different approximants. The main source of this variation is angle interactions that contribute to $E_{tot}$ (as calculated by VASP) but not to our model when we set $\lambda_{72}=\lambda_{144}=0$ in Eq.~(\ref{eq:tileham}).

Note that the values of H, B and S are individually meaningless unless compared with pure elemental energies. For example, the data quoted does not include arbitrary offsets EATOM~\cite{VASP} of each chemical species. However, the difference $2E_t^{B}-E_t^{H}-E_t^{S}=-1.35$ eV {\em is} meaningful because the offsets cancel out (since S$_1$H$_3$ has the same number of atoms of each type as B$_2$H$_2$). Thus a pair of boats is favored over a hexagon-star pair. 

\begin{figure}
\centerline{\epsfig{figure=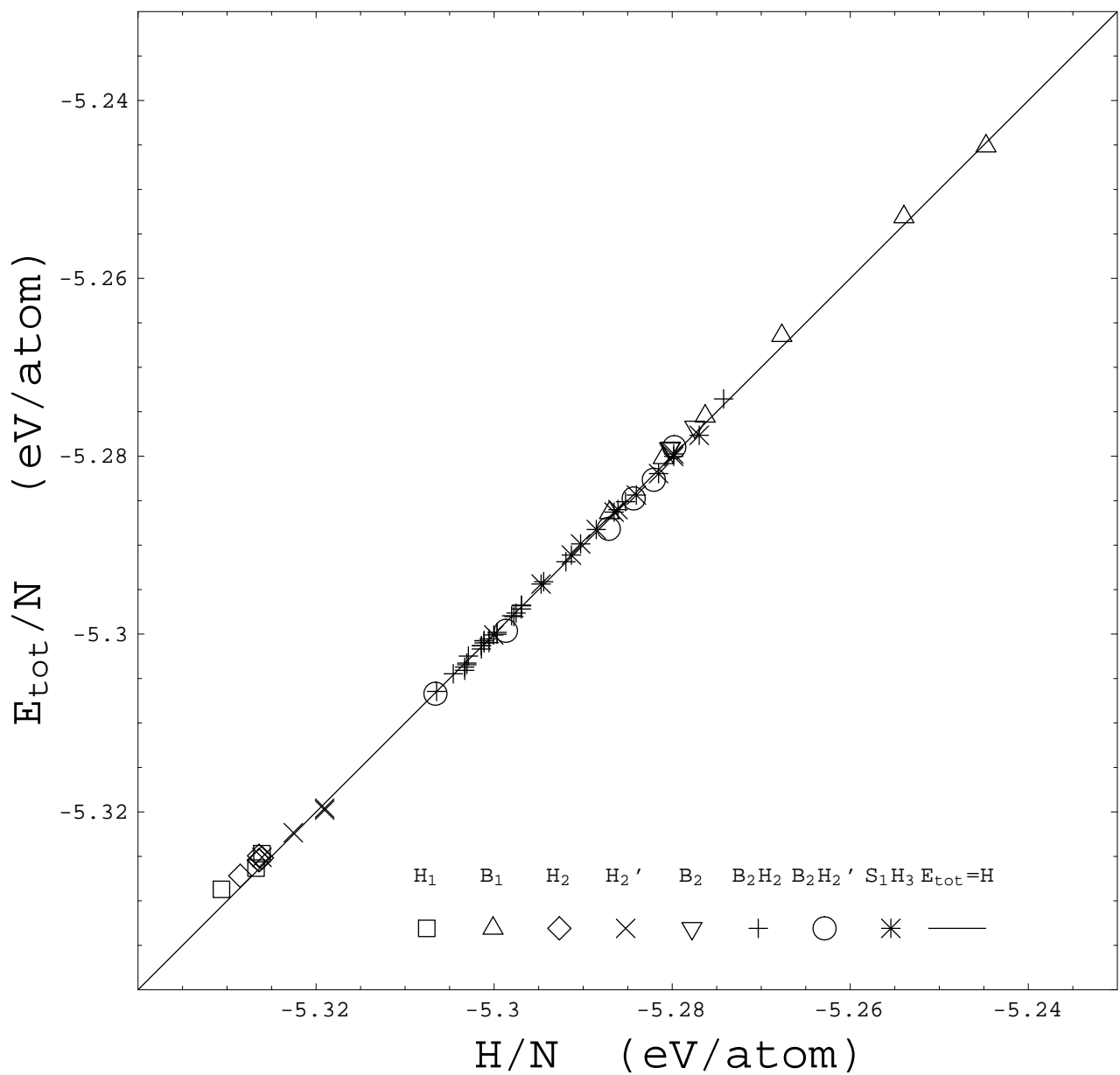,angle=0}}
\vspace{20pt}
\caption{ Including the angle interactions (setting $\lambda_{72}=\lambda_{144}=1$) greatly improves the fitting.}
\label{fig:vasp9}
\end{figure}

The five-parameter fit values are shown in the third column of table~\ref{tab:fitting} where we set $\lambda_{72} = 1$ to include $E_{72}^{\beta}$ and set $\lambda_{144} = 0$ to exclude $E_{144}^{\beta}$. The 72$^\circ$ angles are all internal to the tiles so their energies can't be separated from the tile energies. Only differences in energies can be calculated, such as $E_{72}^{i}-E_{72}^{o}$ and $E_{72}^{m}-E_{72}^{i}$, so we set the lowest energy orientation $E_{72}^o$ equal to zero without any loss of generality. The fourth column of table~\ref{tab:energycosts}~ shows the eight-parameter fit when we set $\lambda_{72} =\lambda_{144} = 1$. In contrast to the number of 72$^\circ$ angles, the number of which is determined entirely by the number of tiles, the number of 144$^\circ$ angles depends on the arrangements of tiles. As a result, we may calculate all three energies $E_{144}^{\beta}$ independently. The fitting is shown in Fig.~\ref{fig:vasp9}. The remaining deviation from the y=x line (standard deviation=0.0013 eV) is due to other effects not included in our model $H$ (Eq.~\ref{eq:tileham}), as well as incomplete convergence or other calculational inaccuracies.

\section{Pair Potentials}
\label{sec:pairpotentials}

The ground state total energy of a system can be expanded in terms of a volume energy and potentials describing n-body (n=2,3,4,...) interactions~\cite{hafner87}. The volume energy is the dominant contribution to cohesive energy. It depends on the composition and density, but not the specific structure. The n-body interactions distinguish between different crystal structures at the same composition and density. It is customary to truncate this n-body series at the pair potentials (n=2), because they are much easier to calculate and use, and because higher order interactions are often weak. Even for transition metals, with their localized d-band, the truncation at pair potentials proved to be practical~\cite{corrections,macdonald}. Pair potentials are functions $V^{\alpha \beta}(r)$ of pair separation $r$ and atom types $\alpha$ and $\beta$.

Many different pair potentials have been used to study this and other quasicrystals. Cockayne and Widom~\cite{CWmodel} proposed mock-ternary potentials extracted from Al-Co pair potentials. Noting that, in Al-Co-Cu, Cu substitutes for an equal combination of Al and Co, they approximated Cu interactions by the average interactions of Al and Co. In addition, the Co-Cu interaction was defined as the average of the Co-Co and Cu-Cu interactions in order to obtain ternary potentials from the AlTM binaries. They adopted Al-Co pair potentials calculated by Phillips et al.~\cite{phillips}. Their discovery of alternation of CoCu pairs atoms on tile edges, and many other details, are all consistent with our VASP results.

Later, more rigorous pair potentials derived by Generalized Pseudopotential Theory (GPT) were developed for Al-Co-Ni and Al-Co-Cu~\cite{corrections,WAM}. The original GPT pair potentials suffered from TM over-binding which is an unphysical attraction between TM atoms at small separations. The strongest over-binding appears in Co-Co pair potentials. We modified the CoCo and NiNi pair potentials at short distances by adding a repulsive term using VASP to get the energy and length scale~\cite{corrections}. The resulting potentials behave really well in simulations~\cite{au8}. The Co-Cu pair potentials were defined as equal to the Ni-Ni pair potentials $V^{\mbox{\scriptsize CoCu}}(r) \equiv V^{\mbox{\scriptsize NiNi}(r)}$. These, in turn, were close to the average of Co-Co and Cu-Cu potentials because Ni resides between Co and Cu in the periodic table. Specifically, $V^{\mbox{\scriptsize NiNi}}(r) \approx \frac{1}{2}(V^{\mbox{\scriptsize CoCo}}(r)+V^{\mbox{\scriptsize CuCu}}(r))$, with biggest error of 0.002 eV at 3.12~\AA~ which is about 15\% error. The other AlCoCu pair potentials were found to be well behaved up to large Cu composition~\cite{WAM}.

We calculated the energies of the approximants using both mock-ternary and modified GPT pair potentials to check how pair potential results compare to VASP. Results of the fitting are summarized in table~\ref{tab:energycosts} for all the methods used. Aside from a difference in energy scale between the modified GPT and the mock-ternary pair potential calculations, they are qualitatively close to each other and to VASP. The order of 144$^\circ$ angle interaction is reversed compared to VASP, but these interactions are very weak.

In table~\ref{tab:energycosts}, we see that $E_{72}^{m} \approx \frac{1}{2}(E_{72}^{i}+E_{72}^{o})$ and $E_{144}^{m} \approx \frac{1}{2}(E_{144}^{i}+E_{144}^{o})$ for all three calculation methods (VASP, mGPT, and mock-ternary). To understand this, note that when two tile edges meet at a vertex, the TM bonds on them are at three different separations from each other. One separation length $r_{i}$ is between the TM positions close to the vertex, $r_{o} $ is between the far positions and $r_{m}$ is the separation between mixed positions. Take the smallest of all, $r_{i}$, as an example and consider pair interactions. In bonds with the  ``i'' configuration, two Co atoms are distance $r_{i}$ from each other. The energy contribution due to this pair is $V^{\mbox{\scriptsize CoCo}}(r_i)$. The same positions are occupied by two Cu atoms in the ``o'' configuration with energy contribution $V^{\mbox{\scriptsize CuCu}}(r_i)$, and by one Co and one Cu atom in the ``m'' configuration with energy $V^{\mbox{\scriptsize CuCo}}(r_i)$. The contribution to the energy difference $E_{72}^{m} -\frac{1}{2}(E_{72}^{i}+E_{72}^{o})$ calculated from these pairs at separations $r_i$ is $V^{\mbox{\scriptsize CuCo}}(r_i)-\frac{1}{2}(V^{\mbox{\scriptsize CoCo}}(r_i)+V^{\mbox{\scriptsize CuCu}}(r_i))$. Similar identities hold at the separations $r_o$ and $r_m$. In mock-ternary pair potentials, these differences of potentials are defined to be zero, suggesting $E_{\theta}^{m}=\frac{1}{2}(E_{\theta}^{i}+E_{\theta}^{o})$ should hold exactly. The small deviations from this identity in the fourth column of table~\ref{tab:energycosts} are due to the fitting procedure. The potential differences are again close to zero for mGPT pair potentials as discussed above, and the small deviations from the energy identity in the third column of table~\ref{tab:energycosts} are also due to the fitting. In VASP, energies are calculated accurately, considering all n-body interactions. An averaging of interactions is not assumed {\em a priori}, but our calculations confirm that averaging {\em is} a good approximation, as shown in the second column of table~\ref{tab:energycosts}.

Another interesting near-degeneracy occurs in a zigzag of 72$^{\circ}$ angles. Such a zigzag runs vertically across Fig.~\ref{fig:h2_mono}. Consider three consecutive bonds in a zigzag. If both outer bonds point in towards, or both point out from, the middle bond, a swap of the middle bond leaves unchanged the total number of ``i'', ``o `` or ``m'' interactions. If one of the outer bonds point in towards the middle bond and the other points out from it, then the swap of the middle bond changes the energy by $2E_{72}^{m}-E_{72}^{i}-E_{72}^{o}$. Because $E_{72}^{m}$ is very close to the average of $E_{72}^{i}$ and $E_{72}^{o}$ (as previously shown) these configurations are again nearly degenerate.

\section{Other Effects}
\label{sec:othereffects}
Chemical ordering of TM atoms on tile edges define edge arrowing in our model. We study chemical ordering here using our H$_2$ approximant (the orthorhombic unit cell in Fig.~\ref{fig:h2_mono}) which contains two horizontal tile edges. When a Cu atom from one horizontal edge is swapped with the Co atom on the other, the resulting edges contain pairs of similar species (one CoCo and one CuCu). The process raises the energy by 0.68 eV/cell. The same swap was studied before with LSMS~\cite{ICQ7} and gave 0.17 eV/cell. Using the pair potentials, TM atoms favor alternation on the tile edges by 0.022 eV/cell for mGPT and 0.079 eV/cell for mock-Ternary. Although the magnitude is not certain, the sign consistently favors Co/Cu ordering.

In Al-Co-Ni, CoCo and NiNi pairs are slightly preferred over CoNi pairs. As a result Al-Co-Ni has no arrow decorations at low temperatures. Cu and Ni are adjacent in the periodic table, but they are notably different in their properties. In an isolated Ni atom, the 3d shell has six electrons and the 4s shell is filled. The partial filling of the d-band strongly influences atomic interactions. The 3d shell in Cu is filled with electrons and the 4s has one electron which makes Cu act more like a simple metal. The d-band of Cu is buried and doesn't participate strongly in interactions. This is why chemical ordering is strong for CoCu pairs but not for CoNi pairs.

An important issue is the position of the symmetry-breaking Al atom inside a hexagon we mentioned in section~\ref{sec:decagonalalcocu}. There are two symmetrically related positions between the two internal Co atoms, and we force the Al atom to take one of these positions as shown in Fig.~\ref{fig:penrose}b (left). If the horizontal edge arrows are parallel to each other, the Al atom prefers to reside in the side closest to the Co atoms by about 0.03 eV. With the off-center Al, we define the decomposition of the hexagon into rhombi such that the symmetry-breaking Al is placed as in Fig.~\ref{fig:penrose}. The position of the internal Al atom inside a hexagon, together with the horizontal tile edge arrows, define a ``direction'' for the hexagon. We noticed that generally hexagons prefer to align parallel to each other in our H$_2$ structure by about 0.01 eV. These effects are very small but are enough to account for some of the discrepancies between $E_{tot}$ and $H$ in our calculations. 

The decomposition of the hexagon into rhombi is lost by placing the Al atom exactly at the center of the hexagon. However, this position is lower in energy by 0.2 eV as calculated by VASP. In the pair potential picture, the central position for the Al atom is preferred by 0.11 eV/cell using mGPT and 0.01 eV/cell using mock-ternary potentials. Depending on the edge decoration, this Al may relax very slightly from the central positions, but this effect is minimal and does not significantly influence the energy. Thus a more realistic model in which the Al atoms are centered should be described even more accurately by our tile Hamiltonian.

One more small effect appears in ``hidden'' 144$^{\circ}$ angles, where two 72$^\circ$ angles share one edge making an extra 144$^\circ$ (Fig.~\ref{fig:b1} has two hidden 144's). The shared edge orientation affects the angle interaction $E_{144}^{\beta}$ of the outer edges. We calculate the difference E$_{144}^o$-E$_{144}^m$ with the shared edge pointing outward and again with it pointing inward. With the shared edge outward pointing, the difference E$_{144}^o$-E$_{144}^m = -0.075$ eV. An inward-pointing middle edge raises the difference by 0.015 eV, so that $E_{144}^{o}-E_{144}^{m} = -0.060$ eV. This effective three-arrow interaction can account for more of the remaining small discrepancies between $E_{tot}$ and $H$.

So far we have examined interactions within the quasiperiodic plane. Now consider perpendicular interactions. Pairs of TM atoms on tile edges are 1.51~\AA~ apart within the quasiperiodic plane and 2.07~\AA~ apart along the perpendicular, periodic direction. The net bond length is 2.56~\AA. The lines connecting them make a zigzag of alternating TM atoms extending along the periodic axes. We turn our attention to atomic order in this direction. Approximant   B$_2$H$_2$ (Fig.~\ref{fig:app132}a) has a horizontal glide plane parallel to the long side of its unit cell that can be exploited for this purpose. We swap one CoCu pair on a horizontal edge (call this pair a) and call the structure (A). Another structure (B) is made from   B$_2$H$_2$ by swapping instead the glide-equivalent image of pair a (call this pair b). These two structures have equal energies by symmetry. Further, we build a 264 atom unit cell by stacking two 132 atom unit cells. It is built once by stacking an A layer over an A layer and another time by stacking a B layer over an A layer. In the AA stacking, TM alternation along the vertical zigzag is conserved. In AB the zigzag sequence is violated along pair a and along pair b. Along each pair the alternation defect includes a CoCo pair and a CuCu pair. The AA and AB structure energies are calculated with a k-point mesh of 2x2x5. The difference $E_{AB}-E_{AA}=0.392$ eV per 264-atom cell.

\section{Discussion}
\label{sec:dis}
We discuss here the implications of our findings on the structure of decagonal AlCoCu. The main result is that now energy can be calculated quickly and accurately for these compounds by adding the relevant terms in the tile Hamiltonian $H$ (Eq.~\ref{eq:tileham}) using parameters obtained in table~\ref{tab:energycosts}. For example, consider the cohesive energy of each tile type. We define a tie-line energy $E_{tie-line}$ to be the energy per atom of the pure element: fcc Al, fcc Cu and spin-polarized hcp Co. The structure energies lie below the tie-line and the difference is the cohesive energy per atom, $E_{coh}$. We calculate $E_{tot}[\mbox{Al}]=-4.17$ eV/atom, $E_{tot}[\mbox{Cu}]=-4.72$ eV/atom and $E_{tot}[\mbox{Co}]=-8.07$ eV/atom, all at the experimental lattice constants. The tile cohesive energies are $E_{coh}[\mbox{H}]=$-7.75 eV , $E_{coh}[\mbox{B}]=$-12.4 eV  and $E_{coh}[\mbox{S}]=$-16.81 eV (using data from our eight-parameter fit). The difference between two boats and a hexagon-star pair is $2 E_{coh}[\mbox{B}]-E_{coh}[\mbox{H}]-E_{coh}[\mbox{S}]=$ -0.24 eV. We can add up the cohesive energies of the tiles to obtain a quick estimate of the cohesive energy of the quasicrystal. For HBS tilings, the ``golden'' ratio H:B:S=$\sqrt{5}\tau$:$\sqrt{5}$:$1$ can be obtained, for example, by removing double-arrow edges from a Penrose tiling~\cite{au8,Henley867}. For such a tiling the cohesive energy is -0.3035 eV/atom. Our results show that stars are disfavored, and a tiling with hexagon and boats is lower in energy. The ratio of H:B in HB tilings is 1:$\tau$ and the cohesive energy is -.3045 eV/atom. 

\begin{figure}
\centerline{\epsfig{figure=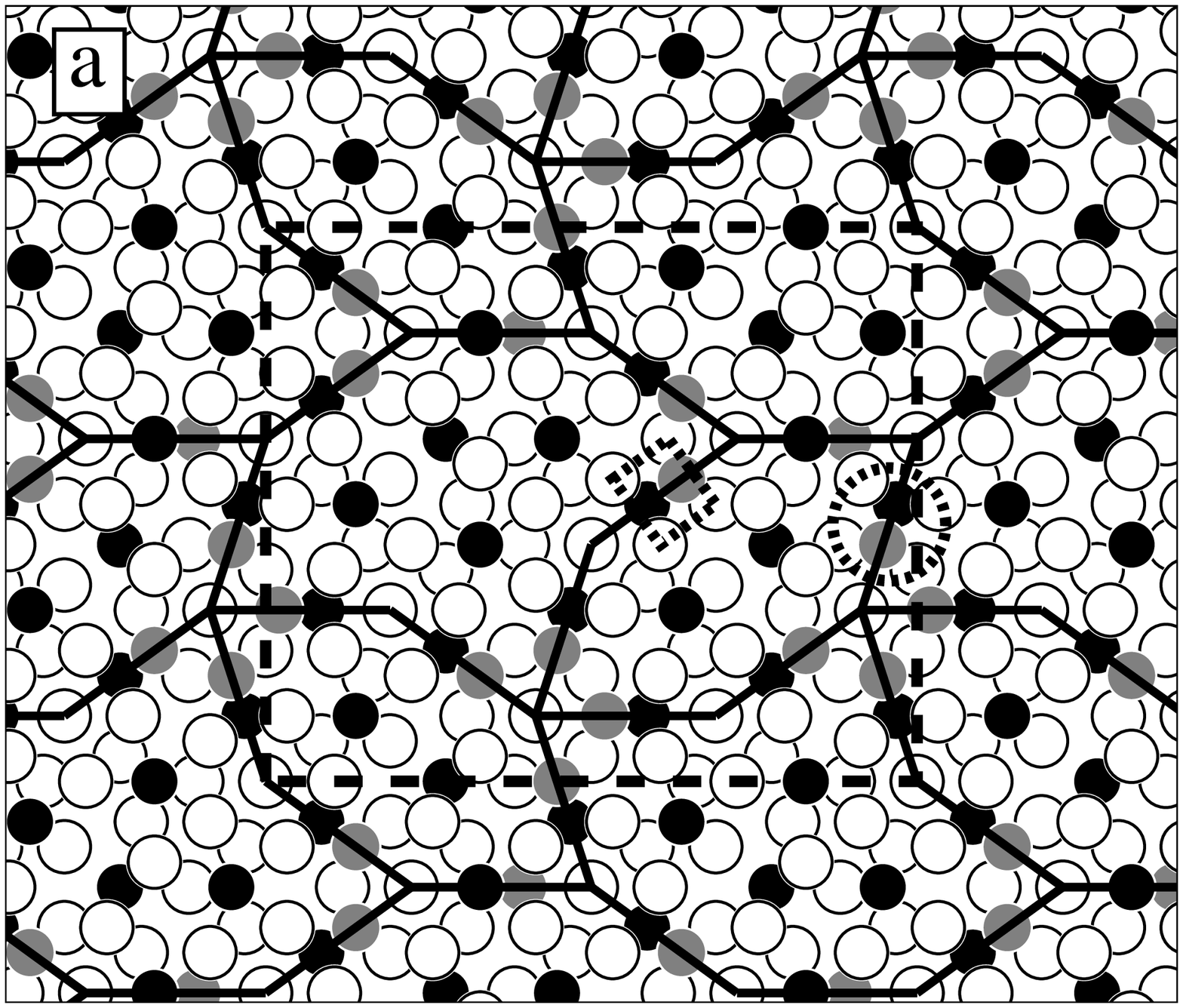,height=3in}}
\centerline{\epsfig{figure=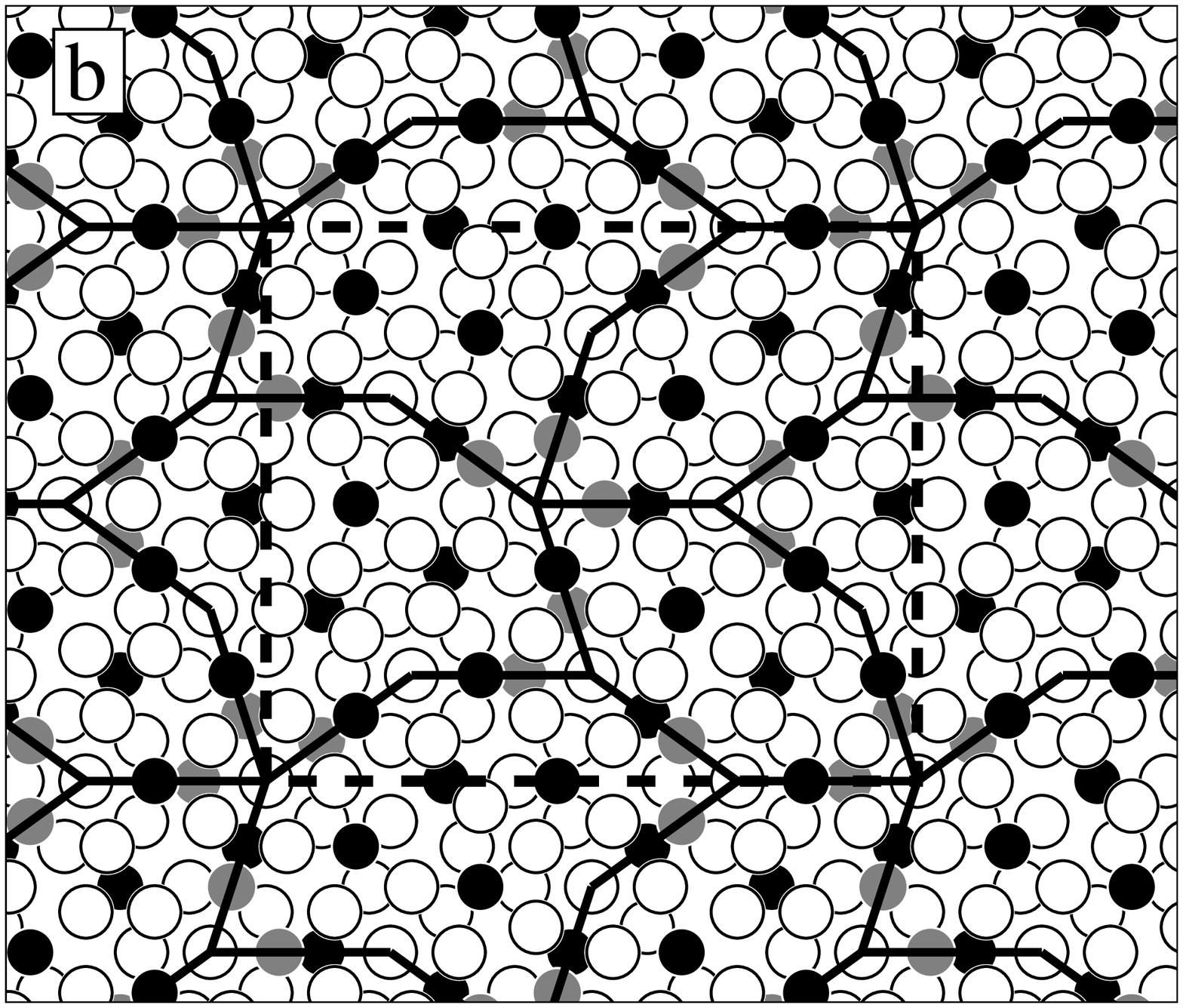,height=3in,angle=-90}}
\centerline{\epsfig{figure=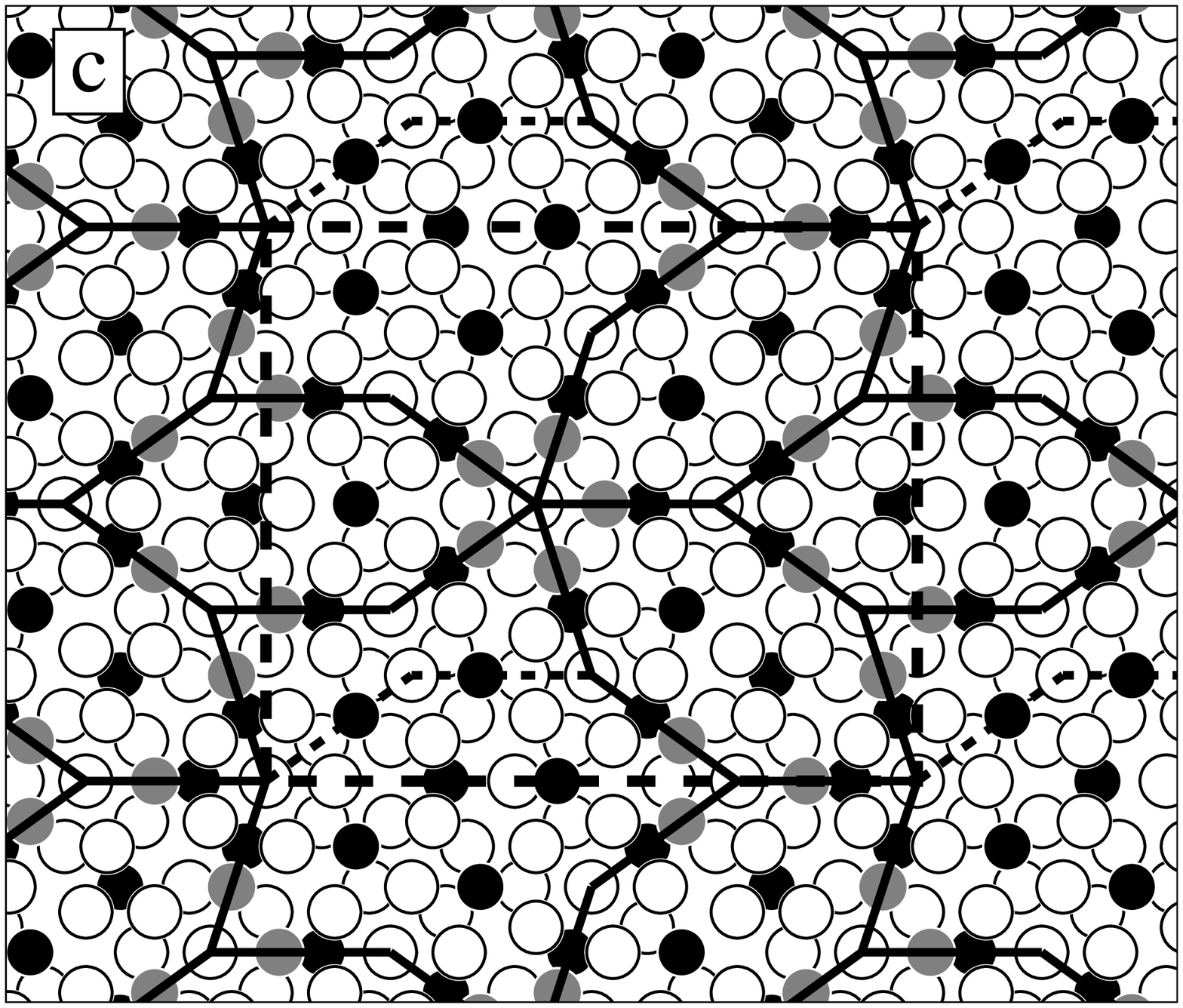,height=3in,angle=-90}}
\caption{Space can be tiled in many ways using HBS tiles. All these approximants (a=B$_2$H$_2$, b=B$_2$H$_2'$ and c=S$_1$H$_3$) have 132 atoms per unit cell. Structures in (b) and (c) differ by a phason flip outlined in (c) with a small dashed line. Bonds surrounded by the dashed square and circle in (a) are bonds that can give information about pure angle interactions.}
\label{fig:app132}
\end{figure}

\begin{figure}
\centerline{\epsfig{figure=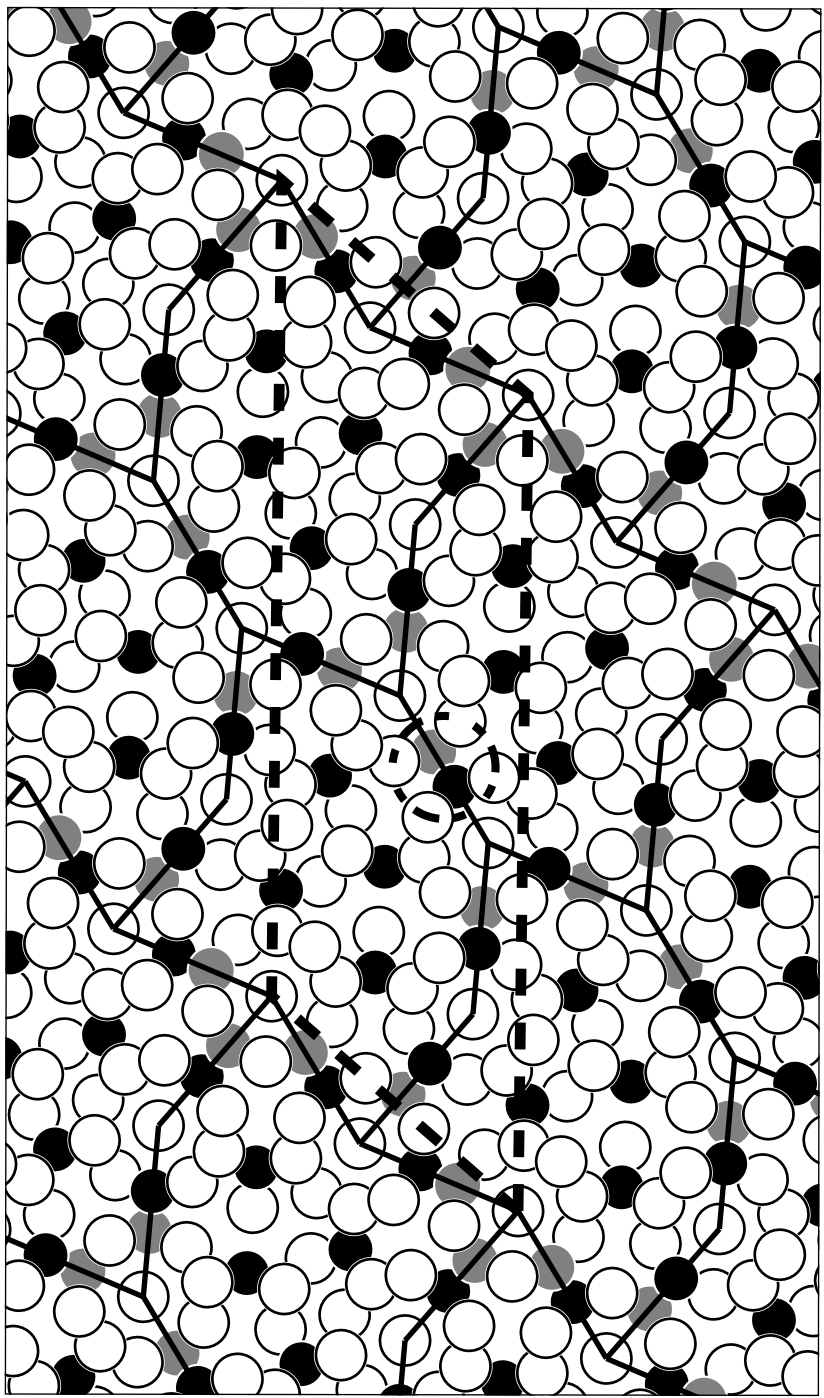,width=4.5in,angle=0}}
\vspace{20pt}
\caption{Two-boat approximant (B$_2$). One of the ``keel'' bonds (circled) in the boats participates only in 144$^\circ$ angles and has a highly symmetric Al environment}
\label{fig:b2}
\end{figure}

Most bonds participate in combinations of 144$^\circ$ and 72$^\circ$ angles. The stronger interactions determine bond arrowing. When a bond is surrounded by a total of four 144$^{\circ}$ angles and no 72$^\circ$ angles, the middle bond is a part of 144$^{\circ}$ zigzag and its decoration doesn't matter. An example of this is circled in Fig.~\ref{fig:b2}. There is only one configuration where a bond orientation {\em is} determined by 144$^{\circ}$ interactions. This is the configuration we used to get pure 144$^{\circ}$ angle effects (see sec.~\ref{sec:abinitiostudy}). These configurations occur occasionally (one is in Fig.~\ref{fig:app132}b), but usually bond orientations are determined primarily by 72$^\circ$ interactions.

Quasicrystals are observed to be stable mainly at high temperatures~\cite{ritsch8}. This can be due to a variety of entropic contributions. Transitions from crystal to quasicrystal phases are reported at about T$\approx$1000K~\cite{transition} or about k$_{B}$T=0.1 eV. At  such temperatures the 144$^{\circ}$ angle interactions are irrelevant because they are small compared to energy fluctuations, and the structure is determined primarily by its tile types and by the 72$^{\circ}$ angle interactions.

Our model expectations are in reasonable agreement with calculated energies, suggesting that we have captured the most important energetic effects. The worst deviation is about 0.1 eV. Out of that we account for 0.03-0.05 eV from the internal Al atom effects on tile edges. The rest can be a collection of long range interactions. We do see these long range effects in some instances. For example, when calculating pure 72$^\circ$ angle interactions using the bond surrounded by a rectangle in Fig.~\ref{fig:app132}a in two different approximants (B$_2$H$_2$ and S$_1$H$_3$). The environments are identical up to about 7~\AA, but a difference of about 0.02 eV in E$_{72}^{m}$-E$_{72}^{o}$ between the two cases shows up.

Pair potential calculations show that they are capable of catching qualitatively the dominant 72$^\circ$ interactions we are investigating with a lot less calculation time.

In our previous paper~\cite{ICQ7}, we reported several results related to edge arrowing calculated using an all-electron multiple scattering method known as LSMS~\cite{Wang985}. Approximants H$_2$ and H$_2'$ were used, with internal Al atoms centered. The swap energy for chemical ordering agrees in sign, but VASP's is four times bigger that LSMS. Other swaps that give 2E$_{72}^m$-E$_{72}^i$-E$_{72}^o$ agree in sign, with a similar factor disagreement in magnitudes.

Further studies might include the effect of TMAl (as opposed to CoCu) arrows on tile edges. The difficulty comes from the fact that such arrowing exist not only on tile edges but also inside the tiles. Phason disorder along the periodic axes is important. So far we studied only Co/Cu disorder along the periodic axis but not tile flips. The system's behavior under relaxation and the preferred relaxed atomic positions are wide areas to explore. Relaxation may alter the quantitative values of our tile Hamiltonian parameters. Finally, the biggest unresolved question is: what type of structure minimizes the value of our tile Hamiltonian?

\acknowledgments
The authors thank C.L. Henley and Y. Wang for useful discussions. IA wishes to thank King Abdul Aziz University (Saudi Arabia) for supporting his study and MW acknowledges support by the National Science Foundation under grant DMR-0111198. We thank the Pittsburgh Supercomputer Center for computer time used for this study.

\begin{table}
\caption{Basic tiles in HBS model and their compositions.}
\vspace{10pt}
\begin{tabular}{|c|l|c|}
\hline
Tile & Composition & Penrose Rhombi\\
\hline
H &  Al$_{17}$Co$_{5}$Cu$_{3}$&2T+F\\
B &  Al$_{29}$Co$_{8}$Cu$_{4}$&T+3F\\
S &  Al$_{41}$Co$_{11}$Cu$_{5}$&5F\\
\hline
\end{tabular}
\label{tab:tiles}
\end{table}

\newpage

\begin{table}
\caption{The approximant tilings we use for our study, their compositions and the number of different decorations (N$_d$) of each of them. The unit cells are either orthorhombic ($a$, $b$ are given) or monoclinic (a, b and $\theta$ are given). All of them have c=4.14~\AA~ in the z-direction. The number of independent k-points is the number on which most of the structures are calculated. To investigate the convergence we go higher for a few structures (see tables \ref{tab:convergenceb1}~and~\ref{tab:convergenceb2h2-1}).}
\vspace{10pt}
\begin{tabular}{|l|r|l|c|c|c|}
\hline
Tiling  &Figure & Composition & $a$, $b$ ($\theta$) & Indep. K-points & N$_d$  \\
\hline
 H$_1$        & \ref{fig:h2_mono} &Al$_{17}$Co$_{5}$Cu$_{3}$     &12.14, 7.50  (72$^\circ$)     &132 &3 \\
 B$_1$        &\ref{fig:b1}       &Al$_{29}$Co$_{8}$Cu$_{4}$     &12.13, 12.13 (108$^\circ$)     &88  &6 \\
 H$_2$        &\ref{fig:h2_mono}  &Al$_{34}$Co$_{10}$Cu$_{6}$    &23.08, 7.56                   &66  &5  \\
 H$_2'$       &\ref{fig:h2_ortho} &Al$_{34}$Co$_{10}$Cu$_{6}$    &14.27, 12.13                   &66  &4  \\
 B$_2$        &\ref{fig:b2}       &Al$_{58}$Co$_{16}$Cu$_{8}$    &12.13, 30.30 (134.5$^\circ$)   &55  &3  \\
 B$_2$H$_2$   &\ref{fig:app132}a  &Al$_{92}$Co$_{26}$Cu$_{14}$   &19.63, 23.08                    &20  &27 \\
 B$_2$H$_2'$  &\ref{fig:app132}b  &Al$_{92}$Co$_{26}$Cu$_{14}$   &19.63, 23.08                   &20  &6  \\
 S$_1$H$_3$   &\ref{fig:app132}c  &Al$_{92}$Co$_{26}$Cu$_{14}$   &19.63, 23.08                    &20  &12 \\
\hline
\end{tabular}
\label{tab:structures}
\end{table}

\begin{table}
\caption{Energies of approximant B$_1$ (Fig.~\ref{fig:b1} and one of its single-swap variants). Our convergence investigation goes through several k-point grids. Nearly isotropic k-point distributions are the 1x1x3 mesh and its multiples. For finer grids, 4x4x11 is more isotropic than 4x4x12.}
\vspace{10pt}
\begin{tabular}{|l|r|c|c|c|}
\hline
K-point grid & Indep. K-points& E$_{B_1}$&E$_{B_1}^{sw}$& $\Delta$E=E$_{B_1}^{sw}$-E$_{B_1}$ \\
\hline
1x1x1  &  1 &-186.83238 &-186.78158 &0.05080 \\
1x1x2  &  1 &-217.02636 &-216.68522 &0.34114 \\
1x1x3  &  2 &-215.95009 &-215.63608 &0.31401 \\
2x2x2  &  4 &-218.18586 &-217.90238 &0.28348 \\
2x2x4  &  8 &-216.75311 &-216.51774 &0.23537 \\
2x2x6  & 12 &-216.74436 &-216.49251 &0.25185 \\
3x3x3  & 14 &-216.59210 &-216.32712 &0.26498 \\
3x3x9  & 41 &-216.74029 &-216.48619 &0.25410 \\
4x4x4  & 32 &-216.77625 &-216.53784 &0.23841 \\
4x4x11*& 88 &-216.74116 &-216.48698 &0.25418 \\
4x4x12 & 96 &-216.74323 &-216.48616 &0.25707 \\
5x5x15 &188 &-216.74291 &-216.48615 &0.25676 \\
\hline
\end{tabular}
\label{tab:convergenceb1}
\end{table}

\begin{table}
\caption{Energies of approximant B$_2$H$_2$ (Fig.~\ref{fig:app132}b and one of its single-swap variants.)}
\vspace{10pt}
\begin{tabular}{|l|r|c|c|c|}
\hline
K-point grid & Indep. K-points& E$_{B_2H_2}$&E$_{B_2H_2}^{sw}$& $\Delta$E=E$_{B_2H_2}^{sw}$-E$_{B_2H_2}$ \\
\hline
 1x1x1&  1 &-606.30421 &-606.22773 &0.07648\\
 1x1x2&  1 &-704.05048 &-703.77620 &0.27428\\
 1x1x4&  2 &-700.48922 &-700.24548 &0.24374\\
 2x2x2&  4 &-704.02670 &-703.77286 &0.25384\\
 1x2x8&  8 &-700.38011 &-700.12094 &0.25917\\
 2x2x4&  8 &-700.73700 &-700.49654 &0.24046\\
 1x2x10& 10&-700.42044 &-700.15409 &0.26635\\
 2x2x8&  16&-700.44533 &-700.18672 &0.25861\\
 2x2x10*&20&-700.45159 &-700.19033 &0.26126\\
\hline
\end{tabular}
\label{tab:convergenceb2h2-1}
\end{table}

\begin{table}
\caption{Fitting our data with different number of parameters. Units are eV. The standard deviation for each set is reported in the last row. Including 72$^\circ$ the 144$^\circ$ angle interactions improved the fits as shown.}
\vspace{10pt}
\begin{tabular}{|c|r|r|r|}
\hline
 Parameters & $\lambda_{72}=\lambda_{144}=0$ & $\lambda_{72}=1,\lambda_{144}=0$ & $\lambda_{72}=\lambda_{144}=1$1 \\
\hline
 E$_t^H$ &-133.17  &-133.15   &-133.15     \\ 
 E$_t^B$ &-216.42  &-216.76   &-216.77     \\
 E$_t^S$ &-298.32  &-300.08   &-300.15     \\
\hline
 2E$_t^B$-E$_t^H$-E$_t^S$ &-1.35  &-0.29 &-0.24       \\
\hline
 E$_{72}^i$ & -      & 0.55   & 0.55       \\
 E$_{72}^m$ & -      & 0.23   & 0.22       \\
 E$_{72}^o$ & -      & 0      & 0          \\
\hline
 E$_{144}^i$ & -     & -      & 0.037      \\
 E$_{144}^m$ & -     & -      &-0.003      \\
 E$_{144}^o$ & -     & -      &-0.034      \\
\hline
 standard deviation       & 0.37  & 0.0026 & 0.0013     \\
\hline
\end{tabular}
\label{tab:fitting}
\end{table}

\begin{table}
\caption{The energy costs of significant parameters in our model obtained from  8-parameter fit}
\vspace{10pt}
\begin{tabular}{|c|r|r|r|}
\hline
 Parameters & Energy (VASP) & Energy (mGPT) & Energy (mock-T)\\
   &  (eV)  & (eV) & (eV)   \\ 
\hline
\hline
 2E$_t^B$-E$_t^H$-E$_t^S$ & -0.24 &    -0.59  & -0.13      \\
\hline
 E$_{72}^i$  & 0.55   & 0.60    & 0.12 \\
 E$_{72}^m$  & 0.23   & 0.31    & 0.05 \\
 E$_{72}^o$  & 0      & 0       & 0 \\
\hline
 E$_{144}^i$ & 0.037  &-0.042   &-0.0079 \\
 E$_{144}^m$ &-0.003  & 0.010   & 0.0001 \\
 E$_{144}^o$ &-0.034  & 0.032   & 0.0080\\
\hline
\end{tabular}
\label{tab:energycosts}
\end{table}

\newpage

\end{document}